# Novel Suboptimal approaches for Hyperparameter Tuning of Deep Neural Network [under the shelf of Optical Communication]

M. A. Amirabadi

Email: m_amirabadi@elec.iust.ac.ir

**Abstract**- Hyperparameter tuning is the main challenge of machine learning (ML) algorithms. Grid search is a popular method in hyperparameter tuning of simple ML algorithms; however, high computational complexity in complex ML algorithms such as Deep Neural Networks (DNN) is the main barrier towards its practical implementation. In this paper, two novel suboptimal grid search methods are presented, which search the grid marginally and alternating. In order to examine these methods, hyperparameter tuning is applied on two different DNN based Optical Communication (OC) systems (Fiber OC, and Free Space Optical (FSO) communication). The hyperparameter tuning of ML algorithms, despite its importance is ignored in ML for OC investigations. In addition, this is the first consideration of both FSO and Fiber OC systems in an ML for OC investigation. Results indicate that despite greatly reducing computation load, favorable performance could be achieved by the proposed methods. In addition, it is shown that the alternating search method has better performance than marginal grid search method. In sum, the proposed structures are cost-effective, and appropriate for real-time applications.

**Keywords**- Hyperparameter tuning, grid search, deep neural network, free space optical communication, fiber optical communication;

1. Introduction

Optoelectronic devices progressed and their cost reduced during the last decades; this caused to attract many considerations towards Optical Communication (OC) system. The OC system has superiorities over Wireless Communication systems including a large un-licensed bandwidth with high security, and simplicity; it is suitable for the last-mile backup/bottle-neck application of the next-generation communication systems [1, 2]. Despite these advantages, OC system has many barriers that could limit its practical applications. For example, the Free Space Optical (FSO) Communication is highly sensitive to atmospheric turbulence, which causes random fluctuations of the received signal intensity [3, 4]; in Fiber OC, nonlinearity plays the role of FSO atmospheric turbulence and is the main limitation of Fiber OC.

Recently, Machine Learning (ML) algorithms used in OC to combat the mentioned effects, and showed favorable performance while using low complexity. ML is a branch of artificial intelligence that help OC systems to learn from input data and improve without additional programming. It focuses on the development of computer programs that can access data and use it for training themselves [5]. However, common ML algorithms were not able to deal with complex OC system or channel models; Deep Learning (DL) (and in its special form Deep Neural Network (DNN)) is a branch of ML that could get very closer this goal [6], and find complex relationships between input and output values [7].

DNN has been widely used in OC systems for various purposes, such as fiber effects mitigation [8], performance monitoring [9], modulation format identification [10,11] and OC network. Almost all of works on DNN for OC are related to Fiber OC, and there is no investigation considering DNN in FSO communication. One of the main ambiguities in working with DNN is selecting (tuning) the DNN hyperparameters. Hyperparameters design the structure of a DNN, and affect the training qualification, therefore, tuning them is very important [12]. The difference between a tuned DNN, and a non-tuned DNN is the difference between an algorithms that could accurately derive a complex relationship, and an algorithm that could not show any improvement.

To the best of the authors' knowledge, there is no investigation in ML for OC considering hyperparameter tuning problem. Accordingly, here some common ML literatures are reviewed to present a background about this subject. Recently, hyperparameter tuning has changed to a new, interesting topic in ML community. This task is challenging, in order to speed up the investigation, some works adjust them empirically or based on prior knowledge from previous literature observations [13]. However, the solutions prepared for hyperparameter tuning are mostly based on trial-and-error, but there are some general solutions including manual search [14], grid search [15], random search [16], as well as Bayesian optimization [17]. Manual search chooses some hyperparameters based on trial and error, and then tune them manually. Grid search first collects all possible parameter combinations, and then tries to find the best combination. Random search collects random parameter combinations and finds the best. Bayesian optimization uses a set of previously evaluated parameters and resulting accuracies to make an assumption about unobserved parameters. Acquisition function using this information suggests the next set of parameters. However, These methods are computationally expensive, because of the requirement to running multiple full training runs that even may increase exponentially while adding hyperparameters to tune (e.g. in a comprehensive examination). So, they are too time consuming, and are not preferable in practical approaches. These methods might be preferable in simple ML algorithms (which have few hyperparameters to be tuned), but, they are not practical in in complex ML algorithms such as DNN (which has many hyperparameters, these methods (exhaustive search)). In these situations, a suboptimal search method, or a method that could automatically find acceptable hyperparameter values in one training run even if the user did not have a strong intuition regarding good values to try for the hyperparameters, would be more practical [18].

Grid search is the most widely used method, because it deals with the trade of between complexity and performance in a better way. This technique searches jointly, and it should be considered that the number of joint values grows exponentially while increasing number of hyperparameters [19]. For dealing with this issue, and with the purpose of reducing the computation of the grid search for hyperparameter tuning of a DNN, this paper presents two novel suboptimal grid search methods. In the first method, marginal search is developed over the grid points, in the sense that at first a grid and an initial point is selected, then one of the hyperparameters is tuned. In this method the whole hyperparameter set is fixed and only one of the parameters is tuned. In the second method, alternating optimization method is used for tuning hyperparameters, in the sense that, at first a grid and an initial point is selected, then one of the hyperparameters is tuned, this point is chosen as the initial point for the next hyperparameter tuning, and then this procedure is repeated until tuning all hyperparameters. In order to examine the accuracy and universality, the proposed methods are applied to two different OC systems (FSO and Fiber OC systems). In these structures, a DNN is used as the receiver, which jointly serves as Equalizer, Detector, and Demodulator. The novelties of this paper include:

1) Presenting a step by step explanation about hyperparameter tuning of a DNN (see section 4).

2) Discussing and solving the problem of hyperparameter tuning in ML for OC applications.

3) Presenting two novel suboptimal grid search hyperparameter tuning methods.

4) Investigating an ML technique applicable for both FSO and Fiber OC systems.

5) Deploying DNN for joint Equalization, Detection, and Demodulation in FSO and Fiber OC systems.

The rest of this article is organized as follows: Sections 2, and 3 present the channel and system models, respectively. Section 4 presents a step by step explanation on hyperparameter tuning of a DNN. Section 5 presents the proposed methods. Section 6 is the results and discussions, and section 7 is the conclusion of this paper.

**2. Channel Model**

In order to show the universality of the proposed method, two completely different applications are considered. These applications are Fiber OC and FSO systems (because the purpose of this paper is to present a new hyperparameter tuning for OC applications). In order to have a fair comparison, all of transceiver parameters of in these systems are assumed to be the same. So, the only difference is the transmission media (channel model), which will be discussed in this section.

**2.1. FSO channel**

Various statistical distributions have been proposed to model atmospheric turbulence of FSO channel, e.g., Exponential-Weibull [20], Generalized Malaga [21], Lognormal [22], Gamma-Gamma [23], and Negative Exponential [24]. Among them, Gamma-Gamma is commonly used and is highly accompanied with the actual results. The probability distribution function of Gamma-Gamma atmospheric turbulence is as follows:

$$f(I) = \frac{2(\alpha\beta)^{\frac{\alpha+\beta}{2}}}{\Gamma(\alpha)\Gamma(\beta)} I^{\frac{\alpha+\beta}{2}-1} K_{\alpha-\beta}(2\sqrt{\alpha\beta I}); \quad I > 0 \qquad (1)$$

where $I$ is the atmospheric turbulence intensity, $\Gamma(.)$ is the well-known gamma function, $K_.(.)$ is Modified Bessel function of the second kind, $\alpha = \left[\exp\left(0.49\sigma_R^2/\left(1 + 1.11\sigma_R^{12/5}\right)^{7/6}\right) - 1\right]^{-1}$ is the shaping parameter, and $\beta = \left[\exp\left(0.51\sigma_R^2/\left(1 + 0.69\sigma_R^{12/5}\right)^{5/6}\right) - 1\right]^{-1}$ is the scaling parameter that characterize the irradiance fluctuation in Gamma-Gamma model, where $\sigma_R^2 = 1.23 c_n^2 k^{7/6} l^{11/6}$, where $k = 2\pi/\lambda$ is wave number, and $l$ is FSO link distance [25].

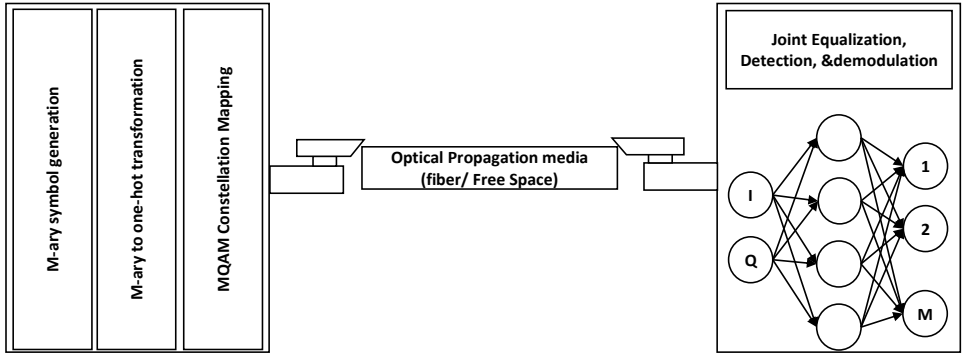

Fig.1. The proposed OC system model

### 2.2. Fiber Channel

Modeling of Fiber optic helps to simpler and more efficient analysis. There are three Fiber optic models available in literature including: the Gaussian Noise model [26], the Non-Linear Interference Noise (NLIN) model [27], and the Extended Gaussian Noise model [28]. The first model assumes statistical independence of the frequency components within all interfering channels and hence models the nonlinear interference as a memory-less AWGN term dependent on the launch power per channel. The second model includes modulation dependent effects and allows more accurate analysis. The last model includes additional cross wavelength interactions, which are only significant in DWDM systems. Accordingly, for the proposed system model in this paper, NLIN model would be sufficient. In this model the propagating signal through Fiber optic is added by an AWGN with zero mean and $\sigma_{NLIN}^2(\cdot) + \sigma_{ASE}^2(\cdot)$ variance, where $\sigma_{ASE}^2$ and $\sigma_{NLIN}^2$ are the variances of the accumulated amplified spontaneous emission (ASE) noise and the NLIN, respectively. $\sigma_{NLIN}^2(\cdot)$ is a function of the optical launch power and moments of the constellation. The NLIN further depends on system specific terms, which are estimated through Monte-Carlo integration and are constant when not altering the system. The Matlab and Python implementation of NLIN could be found in [29] and [30], respectively.

### 3. DNN based OC system model

The proposed DNN based OC system model is shown in Fig.1, in which at first the $N_{sym}$ M-ary symbols are generated at the source, then converted to $N_{sym}$ one-hot vectors of size M (because at the end, the output of the DNN would be an M-ary vector, which should be compared with the transmitted vector). The produced one-hot vectors are then mapped to an MQAM constellation. The mapped signal is transmitted through channel (FSO/Fiber), and added by the receiver AWGN, with zero mean and unit variance.

The receiver is assumed to implement equalization, detection, and demodulation jointly, by use of a DNN. The DNN has two inputs (because the input is complex, but DNN takes real values), and M outputs, $N_{hid}$ hidden layer, and $N_{neu}$ hidden neurons. The activation functions, weights, as well as biases are entitled as $\alpha(.)$, $W$, and $b$, respectively.

The target is to adjust DNN parameters such that, the output M-ary vector of the DNN be the same as the transmitted one-hot vector. Assuming transmitted and detected symbols (M-ary vectors) to be $\boldsymbol{s}$ and $\hat{\boldsymbol{s}}$, respectively, the following loss function should be minimized:

$$L(\boldsymbol{\theta}) = \frac{1}{K}\sum_{k=1}^{K}\left[l^{(k)}(\boldsymbol{s}, \hat{\boldsymbol{s}})\right] \qquad (2)$$

where $\boldsymbol{\theta}$ is the vector containing DNN parameters, $K$ is the batch size, $l(.,.)$ is loss function. The gradients of all layers can be efficiently calculated by the backpropagation algorithm. This loss function could be minimized iteratively using the following formula:

$$\theta^{(j+1)} = \theta^{(j)} - \eta \nabla_\theta \tilde{L}(\theta^{(j)}) \quad (3)$$

where $\eta$ is learning rate, $j$ is the training step iteration, and $\nabla_\theta \tilde{L}(.)$ is the estimate of the gradient. This problem could be solved using the well-known Stochastic Gradient Descent methods [31]. However, in order to do that, first should tune the hyperparameters.

## 4. Hyperparameter tuning of a DNN

Among ML algorithms, DNN has the most number of hyperparameters to be tuned; however, due to lack of investigations, there is a large ambiguity while tuning them. Accordingly, it is necessary to first give a brief but sufficient explanation about the hyperparameters of a standard DNN, and about their tuning. This section presents a step by step explanation, in the sense that it step by step learns that at each step which hyperparameter and how to be tuned. The DNN hyperparameters in the sort include number of epoch and batch size, normalizing input data, selecting the layer type, choosing number of neurons and hidden layers, selecting the activation functions, selecting the loss function, selecting the optimizer, choosing the learning rate and number of iterations.

### 4.1. Selecting number of epochs and batch size

The first step of deploying a DNN algorithm is simply preparing the input data (feature). Terminologies like number epochs and batch size appear while the entering data is large and can't pass through DNN at once. One epoch is when the entire data is passed through the DNN. Since the entering data is large, it should be divided into several smaller batches. Updating DNN parameters in one epoch is not a good idea, and leads to underfitting; accordingly, it should be fed in multiple epochs. As the number of epochs increases, the curve goes from under fitting to good fitting and to overfitting. The number of epochs is different for different datasets, but it is related to how diverse the data is. Batch size is the total number of training samples presented in a single batch.

### 4.2. Normalizing input data

The input data could be either discrete or continuous; the second step in hyperparameter tuning is reducing the variance of input data by scaling. When input data has a limited variation, DNN could better track the changes and derive the relationships, so, normalization helps to accelerate training. The continuous data should be in the range of -1 to 1, 0 to 1 or distributed normally with zero mean and unit variance, and for discrete data, one-hot vector representation could be used. In should be noted that the same normalization method be used for both training and testing data. Also, it is important that if train and test data do not have the same distribution, validation and test data should have the same distribution. There are many ways for normalization available in literature [32], however, the input data of this paper is normalized by itself (one-hot vectors).

### 4.3. Selecting the layer type

The third step in tuning hyperparameters of a DNN is selecting the layer type. There are many layer types available for DNN, which each of them is proper for some specific tasks, and could not be used everywhere. For example, the most popular layer types are Feedforward, Radial Basis Function (for linear data), Multilayer Perceptron (for non-linear data), Convolutional (for imagery data), and Recurrent (for data with memory). In situations that there is no knowledge about the input data the best solution is trial and error. In this paper layer type selection is not investigated, because it has been already selected (DNN, which is an extension of Multilayer Perceptron). However, because there is no note in ML for OC over this subject in this part the famous layer types are reviewed shortly but sufficiently.

Feedforward [33] is one of the simplest types, in which the input data propagates in only one direction through one or more layers. It is used in dealing with high noisy data, in face recognition and computer vision. Radial Basis Function [34] considers the distance of any point relative to the center. It has two layers, in the inner layer, the features are combined with the radial basis function. It is used in power restoration systems. Multilayer Perceptron [35] has three or more fully connected layers. It uses a nonlinear activation function (mainly hyperbolic tangent or logistic). It is used in speech recognition and machine translation. Convolutional NN [36] contains one or more interconnected or pooled convolutional layers. Before passing to the next layer, it applies a convolutional operation on the input. Accordingly, the network can be much deeper but with much fewer parameters. It is used in image and video recognition, natural language processing, recommender systems, semantic parsing, and paraphrase detection. In Recurrent NN [37], except the first layer, other layers have feedback. From each time-step to the next, each node acts as a memory cell and remembers its previous information. It is used in text-to-speech conversion.

**4.4. Number of the neuron and hidden layers**

After selecting the layer type, it's the turn to choose number of layers as well as neurons, which are completely dependent on the input data type. There is no specific formulation for tuning the number of hidden neurons as well as layers. However, there are some empirical rules, e.g. the optimal size of the hidden layer is usually between the size of the input and output layers. In linear data, there is no need for hidden layer. So, usually one hidden layer is sufficient and situations, in which the addition of a second (or third, etc.) hidden layer improves performance are very few. In these situations, the number of neurons is the geometric mean of the neurons in the input and output layers [38].

4.5. **Selecting the activation function**

Table.1. Tensorflow activation functions

| Activation function | Equation |
|---|---|
| Tanh | $f(x) = \dfrac{e^x - e^{-x}}{e^x + e^{-x}}$ |
| Relu | $f(x) = \begin{cases} x & x > 0 \\ 0 & x \leq 0 \end{cases}$ |
| Elu | $f(x) = \begin{cases} x & x > 0 \\ \alpha(e^x - 1) & x \leq 0 \end{cases}$ |
| Selu | $f(x) = \lambda \begin{cases} x & x > 0 \\ \alpha(e^x - 1) & x \leq 0 \end{cases}$ |
| Relu6 | $f(x) = \begin{cases} 6 & 6 < x \\ x & 0 < x \leq 6 \\ 0 & x \leq 0 \end{cases}$ |
| Crelu | $f(x) = \begin{cases} max(0, x) & x > 0 \\ max(0, -x) & x \leq 0 \end{cases}$ |
| Softmax | $f(x_i) = \dfrac{e^{-x_i}}{\sum_j e^{-x_i}}$ |
| Softsign | $f(x) = \dfrac{x}{1 + |x|}$ |
| Softplus | $f(x) = \ln(1 + e^x)$ |

The next step in hyperparameter tuning s specifying the activation function used in neuron, which is somehow related to the task and input data, e.g., Sigmoid and Softmax functions are OK for binary and M-ary classification, respectively, and will lead to faster convergence. In situations that there is no knowledge about the task or data, Relu is a good choice. Actually, Rectifier Linear Unit (Relu) works most of the time as a general approximate. The activation function is either linear or nonlinear; nonlinear activation functions produce output in range (0,1) or (-1, 1), and therefore, could be used for classification whereas linear activation functions produce any output values. This results of this paper are developed in Tensorflow, the available activation function in this environment are collected in Table.1. There is a history behind their generation, which is quoted in the following.

The step function is a threshold based, and activates only when the input is above a certain level. In binary classification it works, but its Achilles Hell is M-ary classification problem, in which multiple neurons are connected. Linear function ($f(x) = cx$) appeared to solve this problem (in situations with more than one firing neuron, max or softmax could be taken). But it has a fixed gradient; in addition, consider a DNN with linear activation function, the each layer output is the next layer input, each firing is based on another linear firing; so, it could be assumed as only one linear firing of the first layer input. Sigmoid function ($f(x) = 1/(1 + exp(-x))$) looks like a smooth and step function, and solved both of previous problems. It has limited analog output, smooth gradient, and nonlinear combinations. However, the input in the range (-2, 2) cause observable change in the output. However, the gradient is small out of this range. Tanh function is a scaled sigmoid function with stronger gradients than sigmoid, and solved this problem somehow.

The Relu is a nonlinear function in range [0, inf) that can be approximated with its combinations. It is a sparse activation function, which only fires a few neurons (almost 50%); it solved the problem of computation of previous functions. However, it has zero gradient for negative inputs, which causes the so-called dying Relu problem, in which the neurons can't respond to the changes. Exponential Linear Unit (ELU) solved this problem by adding a small slope for negative inputs (the slope is defined by a positive constant); furthered, Selu function extended Elu by replacing the linear slope by twisted slope (and an additional constant). These constants are related to the input, e.g., for standard scaled inputs (mean 0, var. 1), the values are $\alpha = 1.6732, \lambda = 1.0507$. Concatenated Relu (Crelu) extended Relu by doing the same in the negative direction.

Softmax function calculates probability distribution function of each target class over all possible target classes. Softsign function is an alternative to hyperbolic tangent. Even though tanh and softsign functions are closely related, tanh converges exponentially whereas softsign converges polynomially. Softplus is an alternative of traditional functions because it is differentiable and its derivative is easy to demonstrate. Sigmoid and tanh outputs have upper and lower limits whereas softplus outputs are in range (0, inf).

### 4.6. Selecting loss function

| Table.2. Tensorflow loss functions |
| --- |
| Softmax cross entropy with logits |
| Sigmoid cross entropy with logits |
| Softmax cross entropy with logits v2 |
| Weighted cross entropy with logits |

After finishing the DNN design, and before start training, should find an appropriate loss function. The same as before, this task is also empirical, dependent on the input data, and so on. For example in M-ary classification, softmax cross entropy, and in binary classification, sigmoid cross entropy are good choices. If the input structure is sparse then sparse softmax cross entropy is preferable, and in situations which one of the classes has a higher weight, weighed cross entropy, which is an extension of sigmoid cross entropy is preferred. Table.2. shows the Tensorflow loss functions. MMSE, and Cross-entropy, are both the famous loss functions; however, Cross-entropy is mostly used in literature. It measures the distance between actual class and predicted value, which is usually a real number between 0 and 1.

Sigmoid cross entropy loss is very similar to Cross-entropy loss function, except that it uses a sigmoid activation function at the last layer. Weighted cross entropy loss is a weighted version of the sigmoid cross entropy loss, which provides a weight on the positive target. Softmax cross entropy loss measures the probability distribution functions by applying a softmax activation function at the last layer. Sparse softmax cross entropy loss is the same as Softmax cross entropy loss, except instead of the target being a probability distribution, it is an index of which category is true. Instead of a sparse all-zero target vector with one value of one, it just passes in the index of which category is the true value.

### 4.7. Selecting optimizer

| Table.3. Tensorflow optimizers |
| --- |
| Adagrad Optimizer |
| Adadelta Optimizer |
| Adam Optimizer |
| Proximal Adagrad Optimizer |
| Gradient Descent Optimizer |
| Proximal Gradient Descent Optimizer |
| Ftrl Optimizer |
| Momentum Optimizer |
| RMS Prop Optimizer |

In order to find the best DNN parameters, the selected loss function in the last section should be minimized, which is done by iterative optimization algorithms. Optimization is a tricky subject, which again depends on the input quality and quantity, model size, and the contents of the weight matrices; again trial and error is the way to determine the best optimizer. The iterative Gradient Descent based formulation for updating DNN parameters is $\theta = \theta - \eta \cdot \nabla J(\theta)$, where $\eta$ is the learning rate, $J(\theta)$ is Loss function, $\nabla J(\theta)$ is the Gradient of Loss function w.r.t parameters $\theta$. The most popular algorithm used in optimizing DNN is Stochastic Gradient Descent (SGD) [39]. However, SGD had high variance oscillations, and couldn't converge properly. This problem was solved by addition of a Momentum term [40], which navigates SGD along the relevant direction and softens the oscillations in irrelevant directions. In the Momentum, the updating function changes to $V(t) = \gamma V(t-1) + \eta \nabla J(\theta)$, and $\theta = \theta - V(t)$, where the $\gamma$ is the momentum term (usually set to 0.9). The momentum term $\gamma$ increases for dimensions whose gradients point in the same directions and reduces updates for dimensions whose gradients change directions.

The momentum is high while reaching the minima (it doesn't slow down at that point), so, it passes the minima. Nestrov Accelerated Gradient solved this problem and prevents going too fast and missing the minima. It takes a big jump according to the previous momentum, then calculates the gradient, makes a correction, and finally updates the parameters. Computing $\theta - \gamma V(t-1)$ gives an approximation of the next position of the parameters. Calculating the gradient w.r.t. the approximate future position of parameter, i.e., $V(t) = \gamma V(t-1) + \eta \nabla J(\theta - \gamma V(t-1))$ gives a look ahead, and finally could update the parameters using $\theta = \theta - V(t)$.

Although this method speeded up the updating, it would be better to apply larger or smaller updates for each individual parameter based on its importance. Adagrad [41] solved this problem by making big and short updates for infrequent and frequent parameters, respectively. Considering $g(t,i)$ to be the loss function gradient w.r.t. to the parameter $\theta(i)$ at time step $t$, the updating formula becomes $\theta(t+1,i) = \theta(t,i) - \eta * g(t,i)/\sqrt{G(t,ii) + \epsilon}$. Actually, it modifies the learning rate at each time step $t$ for every parameter θ(i) based on the past gradients computed for $\theta(i)$. Adagrad doesn't require to know learning rate (a default value of 0.01 would be sufficient); however, its learning rate is always decreasing and decaying. AdaDelta [39] solved this problem by calculating the momentum; it limits the accumulated past gradients to a window with size $w$. The running average $E[g^2](t)$ at time step $t$ then depends only on the previous average and the current gradient. So, the updating formula changes to $E[g^2](t) = \gamma . E[g^2](t-1) + (1-\gamma).g^2(t), \theta(t+1) = \theta(t) - \eta \cdot g(t,i)$.

Adam [40] extended the AdaDelta by calculating momentums for each parameter. In addition to storing an exponentially decaying average of past squared gradients like AdaDelta, Adam also keeps an exponentially decaying average of past gradients $M(t)$, similar to Momentum. The formulas for the first moment (mean) and the second moment (the variance) of the Gradients are $\hat{m}(t) = m(t)/(-\beta_1(t))$, and $\hat{v}(t) = v(t)/(-\beta_2(t))$, where $m(t)$ and $v(t)$ are values of the first and second moment, respectively. The updating formula changes to $\theta(t+1) = \theta(t) - \eta/(\sqrt{\hat{v}(t)} + \epsilon) \times \hat{m}(t)$. RMSprop is similar to Adam it just uses different moving averages but has the same goals. Ftrl-Proximal was developed for ad-click prediction where they had billions of dimensions and hence huge matrices of weights that were very sparse. The main feature here is to keep near zero weights at zero, so calculations can be skipped and optimized.

### 4.8. Learning rate, and number of iterations

The last step in tuning hyperparameters is choosing the learning rate, as well as number of iterations, which is very important. When loss oscillates around a point at the start of training, the learning rate be chosen high. If the loss is decreasing consistently but very slowly, increasing the learning rate is a good idea. Low learning rates not only slow down training but also can even degrade the performance of the model. Large learning rates increase generalization ability. Larger learning rates increase the noise on the stochastic gradient, which acts as an implicit regularizer. Learning rates can take a wide range of values, so gradually adjusting is time-consuming.in addition, results of using learning rates of 0.001 and 0.0011 are not very different. Actually, widely different learning rates should be used to determine the exploring range of learning rates. After finding the optimal range of learning rates, search in smaller grids around the optimal range. Before determining number of iterations, it is required to specify the acceptable model error tolerance. The iterations could be done as much as either reaching a threshold, or failing to make additional progress. In the latter case, ought to adjust hidden layers, consider alternative algorithms, treat data beforehand, or use DL methods.

### 5. Proposed hyperparameter tuning methods

Among hyperparameter tuning methods, grid search is more common and its results are more trustable; however, it is highly dependent on the grid. DNN has much more hyperparameters than the other ML algorithms, so its hyperparameter tuning takes a long time and is not appropriate for time demanding applications. For example, Consider SVM, it has at most two hyperparameters, but DNN at least has 9 hyperparameters (see section 4). In grid search, each of these parameters should get at least 9 points in the grid (because a grid should be wide enough). So, in order to tune hyperparameters of a DNN by grid search method it is required to run 9! = 362880 times the cross validation and see the results, and then decide. Marginally searching the previous scenario would require $9 \times 9 = 81$ computations. Simply developing the grid search marginally rather than jointly might better deserve the tradeoff between time and complexity. How much would be degradation of so much computation reduction? The answer to this question is completely data dependent, however, the results of paper prove that in OC applications, there is slight difference between performances of different hyperparameter sets, and there is no need to develop such investigations. In this section, two novel suboptimal (marginally) grid search algorithms are presented.

### 5.1. First method

At the first step of this method, a grid of hyperparameters and their values should be defined. Then, based on previous knowledge from literature, an initial set (hyperparameter set) is selected. Considering this initial set, one of the hyperparameters is tuned over the defined grid. Again, considering the initial set, another hyperparameter is tuned. This procedure will continue until all of the hyperparameters be tuned. At the second step, the best hyperparameter set of the previous step is selected as the initial point, and the same procedure will be continued. At all of the following steps, the best hyperparameter set is selected as the initial point and the same procedure will be applied. These steps will be repeated until reaching a convergence.

### 5.2. Second method

In the second method, the idea of alternating optimization (which is conducted for multivariate iterative optimization) is used in hyperparameter tuning. At the first step of this method, a grid of hyperparameters and their values should be defined. Then, based on previous knowledge from literature, an initial point (not set!) is selected. Then one of the hyperparameters is tuned, the tuned point is replaced (updated). Then, another hyperparameter is tuned and updated. This process will be repeated until all of the hyperparameters be tuned. The output of the first step is selected as the input of the second step. And the whole procedure of the first step will be repeated. Then, the output of each step will be used as the input of the next step. These steps will be repeated until reaching a convergence.

### 6. Results and discussions

In this section, the proposed DNN hyperparameter tuning methods are implemented in the proposed FSO and Fiber OC systems. The simulations are developed in Tensorflow environment, because it is super helpful for DNN. Tabels.4, and 5 show the parameters of FSO and Fiber OC channels. Table1.6 shows the initial point which is selected based on previous knowledge from literature. Table.7 shows obtained results for hyperparameter tuning of the proposed methods for both of FSO and Fiber OC systems. FSO link is assumed in strong regime of Gamma-Gamma atmospheric turbulence ($\alpha = 4.2, \beta = 1.4$), and $Es/N0 = 0 dB$. The fiber link is assumed to have dispersion, path loss, and nonlinearity. Results indicate that the proposed method of this paper, despite the same computation, could achieve better performance. In addition, it can be seen that both of the proposed methods are dependent on the input data, and hyperparameter tuning is completely different while changing the input data.

In table.7, each of the tuned hyperparameters are bolded and underlined. It should be noted that the in the first method, the best hyperparameter set should be selected, and at the second method, the last hyperparameter set (best accuracy in the last row) should be selected (the best sets of each row are bolded). As can be seen for FSO, the first method results in Symbol Error Rate (SER) of 0.7148, while the second method achieves 0.6992. For Fiber OC, the first method achieves SER of 0.0234, while the second method achieves 0.0117. For simplicity and without loss of generality, only one step results are presented, (because for showing the results of each iteration, a one page length table should be added, and the aim of this paper is not to find the optimum point, it aims to compares the two proposed method results, and their difference is obvious even at the first step).

Tabel.4. FSO channel parameters

| | |
|---|---|
| $\alpha$ | 4.2 |
| $\beta$ | 1.4 |
| Es/N0 [dB] | 0 |

Tabel.5 Fiber OC channel.

| | |
|---|---|
| $C$ | 299792458 |
| $h$ | 6.6261e-34 |
| $D$ | 16.4640 |
| $\beta_2$ | 21 |
| $f_c$ | 1.9341e+14 |
| $\gamma$ [1/W/km] | 1.3 |
| $\alpha$[dB/km] | 0.2 |
| Number of spans | 20 |
| Span length [km] | 100 |
| Pre-dispersion [ps^2] | 0 |
| $P_0$ [dBm] | 2 |
| Baud-rate [GHz] | 32 |
| Channel spacing [GHz] | 50 |
| Second order modulation factor | 1.32 |
| Third order modulation factor | 1.96 |
| EDFA Noise figure | 5 |

Table.6. The initial point of both methods for both system models.

| Hyperparameter | Value |
|---|---|
| Modulation order | 16 |
| Number of layers | 2 |
| Number of hidden neurons | 32 |
| Activation function | Selu |
| Sample Size to Batch Size | 8 |
| Batch Size | 128 |
| Learning Rate | 0.001 |
| Iterations | 250 |
| Loss Function | Softmax cross entropy |
| Optimizer | Adam |

Table.7. Results of proposed methods for FSO system

| Learning rate | 0.00005 | 0.0001 | 0.0005 | 0.001 | 0.005 | 0.01 | 0.05 | 0.1 | 0.5 |
|---|---|---|---|---|---|---|---|---|---|
| method 1-FSO SER | 0.7929 | 0.7578 | 0.7422 | 0.7227 | 0.7617 | **0.7148** | 0.7344 | 0.8242 | 0.9453 |
| method 2-FSO SER | 0.7929 | 0.7578 | 0.7422 | 0.7227 | 0.7617 | **0.7148** | 0.7344 | 0.8242 | 0.9453 |
| method 1-Fiber SER | 0.3477 | 0.0820 | 0.0469 | **0.0429** | 0.0469 | 0.0429 | 0.0469 | 0.0547 | 0.0703 |
| method 2-Fiber SER | 0.3477 | 0.0820 | 0.0469 | **0.0429** | 0.0469 | 0.0429 | 0.0469 | 0.0547 | 0.0703 |
| **# of iteration** | **100** | **200** | **300** | **400** | **500** | **600** | **700** | **800** | **900** |
| method 1-FSO SER | 0.7695 | 0.8320 | 0.7383 | **0.7266** | 0.7734 | 0.7656 | 0.8008 | 0.7539 | 0.7852 |
| method 2-FSO SER | 0.7461 | 0.8203 | 0.7461 | **0.7305** | 0.8008 | 0.7539 | 0.7969 | 0.7773 | 0.8125 |
| method 1-Fiber SER | 0.0508 | 0.0313 | 0.0352 | 0.0273 | 0.0313 | 0.0469 | **0.0195** | 0.0195 | 0.0234 |
| method 2-Fiber SER | 0.0508 | 0.0313 | 0.0352 | 0.0273 | 0.0313 | 0.0469 | **0.0195** | 0.0195 | 0.0234 |
| **# of Layer** | **1** | **2** | **3** | **4** | **5** | **6** | **7** | **8** | **9** |
| method 1-FSO SER | 0.7617 | **0.7227** | 0.7344 | 0.7344 | 0.7539 | 0.7461 | 0.7656 | 0.75 | 0.75 |
| method 2-FSO SER | 0.7422 | **0.7305** | 0.7656 | 0.7344 | 0.7383 | 0.7422 | 0.7344 | 0.7695 | 0.7539 |
| method 1-Fiber SER | 0.0391 | 0.0429 | 0.0469 | 0.0429 | **0.0352** | 0.0391 | 0.0429 | 0.0391 | 0.0429 |
| method 2-Fiber SER | 0.0234 | 0.0196 | 0.0234 | **0.0156** | 0.0273 | 0.0273 | 0.0273 | 0.0273 | 0.0352 |
| **# of Neuron** | **10** | **20** | **30** | **40** | **50** | **60** | **70** | **80** | **90** |
| method 1-FSO SER | 0.7422 | 0.7383 | 0.7305 | **0.7188** | 0.7305 | 0.7266 | 0.7461 | 0.7617 | 0.7422 |
| method 2-FSO SER | 0.7422 | **0.7227** | 0.7344 | 0.7305 | 0.7422 | 0.7305 | 0.7344 | 0.7266 | 0.75 |
| method 1-Fiber SER | 0.04297 | 0.0429 | 0.0469 | 0.0429 | 0.0429 | **0.0391** | 0.0429 | 0.0391 | 0.0391 |
| method 2-Fiber SER | 0.0195 | 0.0156 | **0.0156** | 0.0195 | 0.0195 | 0.0234 | 0.0234 | 0.02734 | 0.0273 |
| **Activation function** | **Relu** | **Crelu** | **Elu** | **Selu** | **Relu6** | **Tanh** | **Softmax** | **Softsign** | **Softplus** |
| method 1-FSO SER | 0.7383 | 0.7461 | 0.7383 | **0.7227** | 0.7305 | 0.7461 | 0.8125 | 0.7422 | 0.7383 |
| method 2-FSO SER | 0.7227 | 0.7266 | 0.7305 | 0.7227 | 0.7422 | 0.75 | **0.7188** | 0.7461 | 0.7422 |
| method 1-Fiber SER | 0.0429 | **0.0391** | 0.0469 | 0.0429 | 0.0469 | 0.0469 | 0.3008 | 0.0391 | 0.0391 |
| method 2-Fiber SER | **0.0156** | 0.0234 | 0.0156 | 0.0156 | 0.0156 | 0.0195 | 0.6133 | 0.0234 | 0.0352 |
| **Optimizer** | **Adam** | **Adadelta** | **Adagrad** | **Ftrl** | **Gradient Descent** | **Proximal Adagrad** | **Proximal Gradient Descent** | **RMS Prop** | **Momentum** |
| method 1-FSO SER | **0.7226** | 0.9375 | 0.8086 | 0.9297 | 0.8086 | 0.8086 | 0.8086 | 0.7539 | 0.7578 |
| method 2-FSO SER | 0.7188 | 0.9648 | 0.9648 | 0.9179 | 0.9648 | 0.9648 | 0.9648 | **0.7070** | 0.9570 |
| method 1-Fiber SER | 0.0429 | 0.9063 | 0.5781 | 0.9336 | 0.6875 | 0.5781 | 0.6875 | **0.0391** | 0.0625 |
| method 2-Fiber SER | 0.0156 | 0.8203 | 0.5078 | 0.9531 | 0.6719 | 0.5078 | 0.6719 | **0.0117** | 0.0195 |
| **Sample size/ Batch size** | **1** | **2** | **3** | **4** | **5** | **6** | **7** | **8** | **9** |
| method 1-FSO SER | 0.8086 | 0.7695 | 0.7890 | 0.7929 | 0.7852 | 0.7539 | 0.7813 | **0.7227** | 0.75 |
| method 2-FSO SER | 0.8281 | 0.7656 | 0.6953 | 0.8164 | 0.7422 | 0.7539 | **0.6992** | 0.7070 | 0.7656 |
| method 1-Fiber SER | 0.0664 | 0.0234 | 0.0469 | **0.0234** | 0.0391 | 0.0313 | 0.0234 | 0.0429 | 0.0234 |
| method 2-Fiber SER | 0.0429 | 0.0195 | 0.0352 | 0.0469 | 0.0352 | 0.0273 | 0.0429 | **0.0117** | 0.0313 |
| **Batch size** | **4\*16** | **8\*16** | **16\*16** | **32\*16** | **64\*16** | **128\*16** | **256\*16** | **512\*16** | **1024\*16** |
| method 1-FSO SER | 0.7812 | 0.7578 | **0.7226** | 0.7597 | 0.786 | 0.7754 | 0.7705 | 0.7631 | 0.7681 |
| method 2-FSO SER | 0.7343 | 0.8203 | **0.6992** | 0.7695 | 0.7705 | 0.7588 | 0.7619 | 0.7667 | 0.7677 |
| method 1-Fiber SER | 0.0313 | 0.0391 | 0.0429 | 0.0273 | **0.0264** | 0.0288 | 0.0376 | 0.0294 | 0.0294 |
| method 2-Fiber SER | 0.03125 | 0.0469 | **0.0117** | 0.0254 | 0.0254 | 0.0327 | 0.0300 | 0.0325 | 0.0283 |
| **Loss function** | **Softmax cross entropy** | **Softmax cross entropy v2** | **Sigmoid cross entropy** | **Weighted cross entropy** | | | | | |
| method 1-FSO SER | 0.7226 | **0.7226** | 0.7734 | 0.7578 | | | | | |
| method 2-FSO SER | 0.6992 | **0.6992** | 0.7421 | 0.7148 | | | | | |
| method 1-Fiber SER | 0.0429 | **0.0429** | 0.0429 | 0.0429 | | | | | |
| method 2-Fiber SER | 0.01172 | **0.0117** | 0.0312 | 0.0273 | | | | | |

7. **Conclusion**

In this paper, two novel suboptimal grid search methods entitled marginal grid search and alternating grid search were presented to solve the problem of high complexity of grid search method for hyperparameter tuning of a DNN. In order to examine, and show universality, these methods were applied on FSO and Fiber OC systems (two different OC systems). Results indicated that although the computational load of these methods is greatly reduced compared with joint grid search method, the performance would not be affected so much. In addition, it was shown that the alternating grid search method, despite the same computation, has better performance than the proposed marginal grid search method.